\documentclass[twocolumn,showpacs,amsmath,amssymb,pre]{revtex4}
\usepackage{graphicx,color}
\begin{document}
\title{The glass transition of two-dimensional binary 
soft disk mixtures with large size ratios} 
\author{Rei Kurita}
\altaffiliation{Current address: Institute of Industrial Science,
University of Tokyo, 4-6-1 Komaba, Meguro-ku, Tokyo, 153-8505,
Japan}
\author{Eric R. Weeks}
\affiliation{Department of Physics, Emory University, Atlanta, GA 30322 U.S.A}
\date{\today}

\begin{abstract} 
We simulate binary soft disk systems in two dimensions,
and investigate how the dynamics slow as the area fraction
is increased toward the glass transition.  The ``fragility''
quantifies how sensitively the relaxation time scale depends
on the area fraction, and the fragility strongly depends on the
composition of the mixture.  We confirm prior results for mixtures
of particles with similar sizes, where the ability to form small
crystalline regions correlates with fragility.  However, for
mixtures with particle size ratios above 1.4, we find that the fragility
is not correlated with structural ordering, but rather with the
spatial distribution of large particles. The large particles 
have slower motion than the small particles, and act as confining
``walls'' which slow the motion of nearby small particles.  The
rearrangement of these confining structures governs the lifetime
of dynamical heterogeneity, that is, how long local regions exhibit
anomalously fast or slow behavior.
The strength of the confinement effect is correlated with the
fragility and also influences the aging behavior of glassy systems.
\end{abstract}
\maketitle


\section{Introduction}

Many liquids can form glasses if they are cooled rapidly, and glassy
materials have technological applications such as optical fibers and
plastics \cite{Debenedetti,Angell,Angell2,Ediger,Kob}.  The origin
of the glass transition is still unclear, despite the scientific
and technological interests.  Much work has examined the dynamical
properties of materials near the glass transition.  Those studies
revealed several important features of supercooled liquids and
glasses.  For example, upon approaching the glass transition,
the structural relaxation time ($\tau_\alpha$) increases by
several orders of magnitude without a corresponding growing static
correlation length \cite{ernst91,vanblaaderen95}.  The rate of
this increase of $\tau_\alpha$ is called ``fragility'' and depends
on the material \cite{Debenedetti,Angell, Angell2,Ediger,Kob}.
For fragile glass-formers, $\tau_\alpha$ steeply increases for a
small decrease in temperature, while for ``strong'' glass-formers,
the increase of $\tau_\alpha$ requires a larger decrease in
temperature.  The relaxation time is related to the viscosity,
and thus the fragility is an important factor in the ease of
processing glass-forming materials.  Typically it is desirable
to mold a glass-forming material with the viscosity held within
a certain range; for fragile materials, this may correspond to a
restrictively narrow range of temperature.

Another common observation of materials close to the glass
transition is that they often have a broad distribution of
local mobility: some regions in a sample relax much faster
than other regions.  Eventually, molecules in those regions
exchange their dynamics, that is, fast regions become slow
and vice versa.  This is termed ``dynamical heterogeneity''
\cite{kob97,donati98,Ediger2,Harrowell,Eric1,Berthier,Yamamoto}.
It is known that the characteristic lifetime of these dynamically
heterogeneous regions $\tau_{DH}$ is longer than $\tau_{\alpha}$
and the strong divergence of $\tau_{DH}$ near a glass transition
temperature suggests that heterogeneous dynamics are relevant for
understanding the glass transition \cite{Ediger}.

A third common feature of materials in the glassy state is that
their properties evolve with time, such as the diffusivity of
molecules or dielectric susceptibility.  This phenomenon is termed
``aging'' \cite{Angell3,Hodge,Megen,courtland03,Gianguido,Jenn}.  Despite the
changing properties, no clear structural changes have been seen;
this can be true even if, for example, diffusivity slows by several
orders of magnitude \cite{Gianguido,Jenn}.

While all of these properties have been known for some time
and carefully characterized by experiments, the origins of
many of them are unclear.  To understand the origins, numerical
simulations with simple intermolecular interactions are used to
study factors controlling the dynamics.  These model systems are
useful as clear understanding is hindered by the complexity of
real materials \cite{Onuki,KAT,Sun,Coslovich,Tarjus}.  For the
fragility, some simulations show that liquids become less fragile
when the polydispersity increases (or a larger size ratio for
binary mixtures is used) \cite{KAT,Sun,Coslovich}.  This has
been explained as due to ordering of the sample, which becomes
frustrated in polydisperse samples. For example, in two-dimensional
(2D) systems, small regions with hexagonal order can form which
correspond with slower dynamics (larger values of $\tau_\alpha$),
and thus increasing the polydispersity frustrates formation of
these ordered regions and diminishes the fragility \cite{KAT}.
Furthermore, it is reported that particle mobility in those ordered
regions are slower than that in randomly structured regions and
it suggests that dynamical heterogeneity is also influenced by
local structure
\cite{KAT,Tarjus,weeks02,harrowell04,widmercooper05,widmercooper08,conrad05,matharoo06,berthier07,appignanesi09}.

Those simulations used the polydispersity as a small perturbation
frustrating the ordering, that is, with either a small
polydispersity or a binary system with the particle size ratio
close to 1.  However, dynamics in those situations are quite
different from highly polydisperse samples.  Binary mixtures
have two control parameters: the size ratio and the volume
fraction ratio of the two components.  These lead to complex phase
diagrams \cite{Imhof,dinsmore95} and potentially emergent dynamical
properties such as an effective depletion interaction between the
large particles in binary hard sphere systems \cite{AO,crocker99}.
Our interest is in dense amorphous phases at intermediate size
ratios.

In this article, we simulate the dynamics of binary soft disk
mixture systems with large size ratios and find that the glass
transition in large size ratio binary systems can be quite different
from that of systems with smaller size differences.  We study the
fragility of binary soft disk mixtures with various size ratios
and area fraction ratios.  Local ordering is less significant for
large size ratio systems.  Instead, we see a molecular crowding
effect from the large particles \cite{moreno06,voigtmann09}.  Our data
suggest that the large particles act as confining walls for the
smaller particles, and that confinement effects increase the
fragility of such systems by suppressing dynamical heterogeneity.
We also investigate aging in large size ratio systems, and again
seen an influence of confinement effects due to the large particles.
Overall, our results suggest that confinement effects are crucial to
describing the dynamics of the glass transition in binary mixtures
with large size ratios between the two components.

\section{Method}

We perform two-dimensional Brownian dynamics simulations for binary
mixtures composed of large ($L$) and small ($S$) soft particles.
These simulations are meant to mimic the colloidal glass transition.
For the colloidal glass transition, the key control parameter
is the volume fraction \cite{vanblaaderen95,Eric1,pusey86}, and
so in our simulations the chief control parameter is the area
fraction $\phi$.  The results of Brownian Dynamics simulations
are similar to those of molecular dynamics simulations in dense
systems \cite{KAT,Coslovich,gleim98,szamel04,tokuyama07}.  
The particles interact via the
purely repulsive Weeks-Chandler-Andersen potential \cite{WCA};
$U_{ij} = 4\epsilon [(\sigma_{ij}/r)^{12}-(\sigma_{ij}/r)^6+1/4]$
for $r<2^{1/6}\sigma_{ij}$, otherwise $U_{ij}=0$, where
$\sigma_{ij}=(\sigma_i+\sigma_j)/2$ and $i, j  \in \{L, S\}$.
We fix $k_BT/\epsilon=0.04$, so the total area fraction $\phi$
is our control parameter to approach the glass transition.
The mass ratio is $m_L/m_S$=$(\sigma_L/\sigma_S)^2$ and the length
is normalized by $\sigma_S$.  The total number of particles is
$N=N_L+N_S=1024$ (or 4096) where $N_L$ and $N_S$ are the number
of large and small species, respectively.  We generate initial
conditions by simulating at $\phi = 0.10$ for a long time and
expanding the disk sizes to increase $\phi$.  We confirm that our
results do not show initial condition dependence and our results
are also independent on time below $\phi = 0.66$, that is, there
is no aging observed.  Thus, we consider our binary systems are
well mixed and ergodic for $\phi < 0.66$.

\begin{figure}
\begin{center}
\includegraphics[width=8cm]{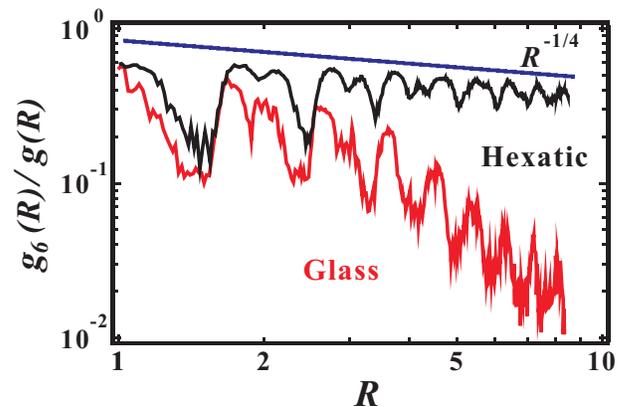}
\end{center}
\caption{(Color online.) The orientation pair correlation as a
function of distance.  The straight line corresponds to $R^{-1/4}$,
the behavior expected for hexatic phases.  A sample with size ratio
$\sigma_L/\sigma_S$=1.2 and area fraction ratio $\phi_L/\phi_S$ =
0.5 at $\phi$ = 0.66 does not have hexatic order since $g_6(R)/g(R)$
decays faster than the power of $-1/4$ (the lower curve, marked
``Glass'').  In contrast, the upper curve (marked ``Hexatic'') shows
the correlation for $\sigma_L/\sigma_S$=1.15, $\phi_L/\phi_S$ =
0.5, and $\phi$ = 0.66, which has long range orientational ordering.
}
\label{hexatic}
\end{figure}

For certain area fractions for a 2D monodisperse sample, the
hexatic order phase can be found, which has orientational
order but no long-range positional order \cite{Halperin}.
For binary samples with large size ratios, there is apparently
little ordering of either type, while for size ratios close to 1,
hexatic phases are still subtle to verify.  Hence, we calculate
an orientation pair correlation function $g_6(R)/g(R)$ where
$R$ is a normalized distance from a particle.  $g_6(R)/g(R)$
decays with the power of $R^{-1/4}$ in a hexatic phase for a
monodisperse sample \cite{Halperin}.  Thus, we classify our
samples as liquid (or glassy) when the power of the decay is
faster than $R^{-1/4}$ [Fig.~\ref{hexatic}].  We find no hexatic
phases for any area fractions $\phi$ for systems with size ratios
$\sigma_L/\sigma_S$=1.2, 1.25, 1.5, 1.75, 2, 2.5 and 3 and area
fraction ratios $\phi_L/\phi_S$=$N_L \sigma_L^2/N_S \sigma_S^2$=0.5,
0.75, 1, 1.5 and 2; these states are indicated by the circles
in Fig.~\ref{contourfig}(a).  We confirm that the decay rates of
$g_6(R)/g(R)$ in those simulation conditions are quicker than
$R^{1/4}$ and thus we are studying liquids (or glasses at higher
$\phi$).  We note that it is known that several crystalline structures
exist in three-dimensional (3D) binary sphere suspensions
\cite{Imhof,kikuchi07}.  In none of our samples do we observe large
hexagonally ordered patches for our simulations, and the small
patches that sometimes appear are transient.  This is discussed
further below.

When the size ratio is large, a depletion force should be generated
\cite{AO}.  This has been studied before in dilute suspensions
of large and small spheres, and manifests itself as an effective
attractive force between the large spheres \cite{crocker99}.
We cannot rule out the existence of the depletion force in
our simulations, which may affect the structure of the large
particles, although at large area fractions the depletion force
may be less relevant.  Indeed, we observe pairs of adjacent large
particles which neighbors for long periods of time, but these
neighbors separate eventually (see Movie~S1, EPAPS materials to
be submitted), and in general all of the dynamics are slow in
our glassy experiments.  It's unclear if depletion introduces
still slower dynamics, or if this is merely part of the overall
slow behavior.  At least, we note that our systems are quite
different from gels with strong attractive force.

\section{Results}

\subsection{Fragility:  behavior as $\phi$ increases} \label{sec:fra}

\begin{figure}
\begin{center}
\includegraphics[width=8cm]{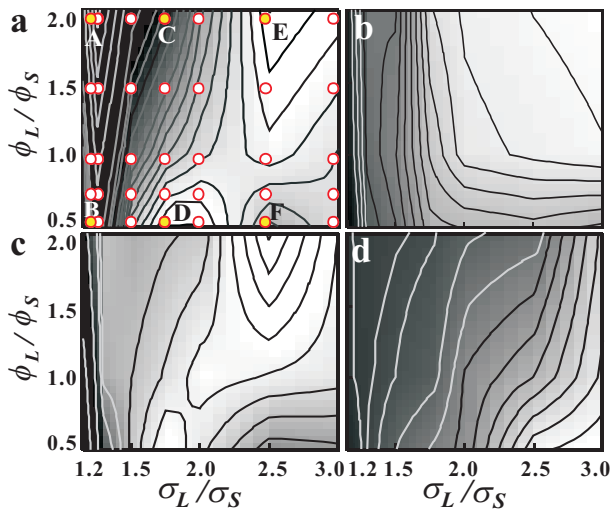}
\end{center}
\caption{(color online). 
The contour plot of (a) fragility index $D$, (b) the growth rate of
hexagonal order $\partial \langle \psi_6 \rangle/\partial \phi$,
(c) the growth rate of dynamical correlation length $\partial
\xi/\partial \phi$ and (d) the dynamical correlation around
a large particle $\Theta$ in a plane of ($\sigma_L/\sigma_S$,
$\phi_L/\phi_S$).  All those figures except (a) are obtained
at $\phi$=0.66.  The shading in each figure is darker when each
value is small, thus, the darker regions correspond to (a)
fragile liquids, (b) low growth of hexagonal order, (c) low growth
of dynamical correlation length and (d) minimal correlations of
motion between a large particle and its neighbors.
The circles in (a) are the simulated
points and the specific states A, B, C, D, E and F are labeled.
}
\label{contourfig}
\end{figure}

\begin{figure}
\begin{center}
\includegraphics[width=8cm]{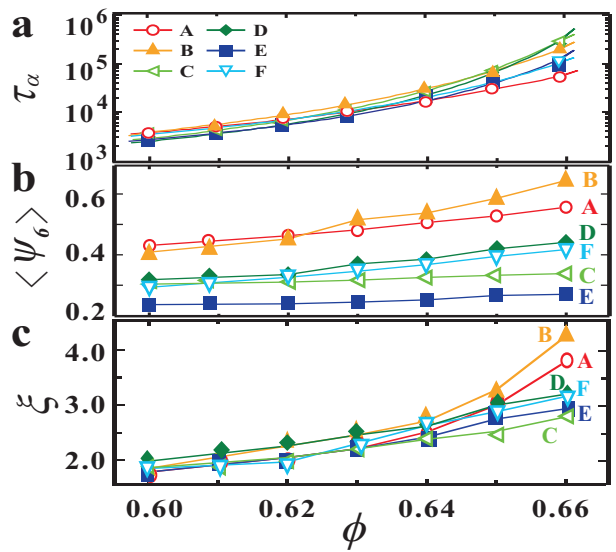}
\end{center}
\caption{(Color online.) The area fraction 
$\phi$ dependence of (a) the relaxation time
$\tau_\alpha$, (b) average hexagonal ordering $\langle \psi_6 \rangle$
and (c) the dynamic correlation length $\xi$ at states A, B, C, D, E
and F indicated in Fig.~\ref{contourfig}(a).  The solid lines in (a)
are fitting lines with the Vogel-Fulcher function at each state. 
Filled and open
symbols correspond to fragile liquids and less fragile liquids,
respectively.
}
\label{phidependence}
\end{figure}

We obtain the relaxation time $\tau_\alpha$ from the self-part of
the intermediate scattering function for all particles, which is
given by $F(k,t)$=$\frac{1}{N} \sum_{j} \langle \exp{i \vec{k}\cdot
[\vec{r}_j(t)-\vec{r}_j(0)]} \rangle,$ where $\vec{r}_j$
is the position vector of particle $j$, $\langle\rangle$
indicates a time average and $\vec{k}$ is the wave vector.
$\tau_\alpha$ is determined when $F(k_p, \tau_\alpha)$=$1/e$
where $k_p$ corresponds to the wave number of the first peak of
the structure factor.  The $\phi$ dependence of $\tau_\alpha$
is well fitted by Vogel-Fulcher function substituting $\phi$
for $1/T$: $\tau_\alpha$=$\tau_0\exp[D\phi/(\phi_0-\phi)]$, where
$D$ is the fragility index and $\phi_0$ is the area fraction
of the ideal glass transition [see Fig.~\ref{phidependence}(a)]
\cite{Angell2}.  Fragile liquids have smaller values of $D$.
For example, $D\sim 4$ for triphenyl phosphite which is one
of the most fragile liquids, while for the less fragile liquid
butyronitrile, $D \sim 30$ \cite{Bohmer2}.  For our simulations,
$0.4 \lesssim D \lesssim 1.0$, smaller than the molecular liquids.
The difference may be due to using the density as the control
parameter rather than temperature.  For comparison, we examined
the data of Refs.~\cite{brambilla09,elmasri09} which used light
scattering to study the colloidal glass transition as a function of
volume fraction.  From their data, we find $D = 0.497 \pm 0.002$.
Another example is that the fragility index of glycerol in an isothermal
experiment is smaller than that measured in an isobaric experiment
\cite{Cook}.  We expect our observed qualitative dependence of
the fragility on the system parameters to still be revealing.

It is worth noting that for our data, we also compute the value
of $D$ by calculating $F(k,t)$ for only small particles (or
large particles) and we obtain almost the same values for $D$
and $\phi_0$.  We do not see separate glass transitions for the
two particle species, which has been seen in simulations with large
size ratios and equal particle numbers (thus volume fraction ratios
$\phi_L/\phi_S$ much larger than ours) \cite{moreno06,voigtmann09}.
In such cases, the large particles do not seem to interact as
directly with the small particles but rather have their own glass
transition, and then the small particles move in the interstices
between the large particles.  In our simulations, the volume
fraction ratios are somewhat close to 1, and so the large
particles always ``see'' the small particles and the alpha
relaxation time scales for both particle species have similar
$\phi$ dependence.

Figure \ref{contourfig}(a) shows the contour plot of the
fragility $D$ in a ($\sigma_L/\sigma_S$, $\phi_L/\phi_S$) plane.
The circles indicate the states simulated.  This figure shows that
the fragility is a non-monotonic function of the size ratio or the
area fraction ratio.  For example, for $\phi_L / \phi_S = 2.0$,
the fragility has values $D=1.05, 0.82, 0.45$ for states A, C,
and E, but then increases slightly to 0.49 for the state to the
right of E. Likewise, for $\phi_L/\phi_S = 0.5$, the fragility
behaves non-monotonically with increasing size ratio: $D=0.64,
0.48, 0.77$ for states B, D, and F.  Comparing states A and B, or
C and D, suggests that increasing the number of large particles
increases the fragility index $D$, but states E and F
disprove this trend.

We find that fragile liquids (small $D$) have small $\phi_0$,
the area fraction where $\tau_{\alpha}$ appears to diverge.
The relationship between fragility and the divergence point of
$\tau_\alpha$ is observed at glass-forming liquids \cite{Tanaka,KT2}
and our results are consistent with them.  On the other hand, the
existence of $\phi_0$ for molecular liquids is still discussed
and not well established \cite{Nm,Phi0}.  We are not sure what
determines $\phi_0$ and why fragility is related to the divergence
point.  Below, we focus on microscopic properties such as particle
mobility and local arrangement.

Prior work observed that
two-dimensional systems can form small
hexagonally ordered regions, and the mobility of particles
is diminished within these regions.  More fragile liquids are
observed to have large growth rates of the size of these regions
with respect to $\phi$
\cite{KAT,Sun,Coslovich}.  To check this we
study the $\phi$ dependence of hexagonal order for our samples.
We use the local hexatic order parameter described as $\psi_6^j =
\frac{1}{n_j}\mid\sum^{n_j}_{m=1} e^{i6\theta_m^j}\mid$ where $n_j$
is the number of nearest neighbors for particle $j$ and $\theta_m^j$
is the angle of the relative vector $\vec{r}_m-\vec{r}_j$
with respect to the $x$ axis.  $\psi_6^j$=1 means that a hexagonal
arrangement is formed around particle $j$, while $\psi_6^j$=0
corresponds to a non-hexagonal arrangement.  We then consider
$\langle \psi_6 \rangle$, the time and particle average of
$\psi_6^j$.  We compare the $\phi$ dependence of $\langle \psi_6
\rangle$ with that of $\tau_\alpha$ [Fig.~\ref{phidependence}(a) and
(b)].  Both particle sizes are similar at states A and B, $\sigma_L/\sigma_S
= 1.2$, and here both $\tau_\alpha$ and $\langle \psi_6 \rangle$
increase dramatically as $\phi$ increases, suggesting they are
indeed related as seen in prior work.  However, for states C and E,
$\langle \psi_6 \rangle$ stays nearly constant with increasing
$\phi$, while $\tau_\alpha$ grows rapidly.  
The system slows without significant hexagonal ordering.

\begin{figure}
\begin{center}
\includegraphics[width=8cm]{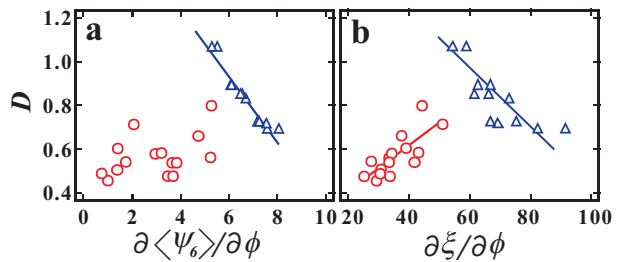}
\end{center}
\caption{(Color online.)  The fragility index $D$ as a function
of (a) $\partial \langle \psi_6 \rangle/\partial \phi$ and (b)
$\partial \xi/\partial \phi$ at $\phi$=0.66. The triangles correspond
to systems where size ratio is close to 1 ($\sigma_L/\sigma_S<1.4$)
and the circles
correspond to large size ratios ($>1.4$).  The lines are added as
guides to the eye.
}
\label{fragilityfig}
\end{figure}

Next, we compare the fragility with the growth rate of hexagonal
ordering $\partial \langle \psi_6 \rangle/\partial \phi$ since
fragility corresponds to $\partial \log\tau_\alpha /
\partial \phi$.  (We calculate all derivatives
of a quantity $X$ with respect to $\phi$ as
$\partial X / \partial \phi = 
[X (\phi ) - X (\phi - \Delta \phi) ]/\Delta \phi$
where we choose $\Delta \phi=0.01$.  In the results below,
we compare the behavior of the samples at a fixed $\phi=0.66$.
However, our results show similar trends when compare samples at
fixed $\tau_\alpha$.)  Figure \ref{contourfig}(b) shows the contour
plot of $\partial \langle \psi_6 \rangle/\partial \phi$ at $\phi
= 0.66$ in a plane of ($\sigma_L/\sigma_S$, $\phi_L/\phi_S$).
The growth of hexagonal order is small at large $\sigma_L/\sigma_S$
and $\phi_L/\phi_S$ (upper right region).  This behavior is expected
since hexagonal ordering should be frustrated with increasing
$\sigma_L/\sigma_S$.  Figure \ref{fragilityfig}(a) shows $D$ as a
function of $\partial \langle \psi_6 \rangle/\partial \phi$.
We observe two distinct behaviors.  For similar particle size
systems ($\sigma_L/\sigma_S<1.4$, triangles), more fragile liquids
(smaller $D$) have a larger dependence of hexagonal order on
$\phi$, in agreement with prior work \cite{KAT}.  In contrast,
large size ratio systems ($\sigma_L/\sigma_S>1.4$, circles) show
less correlation between the growth of hexagonal order and
the fragility index.

\begin{figure} 
\begin{center}
\includegraphics[width=8cm]{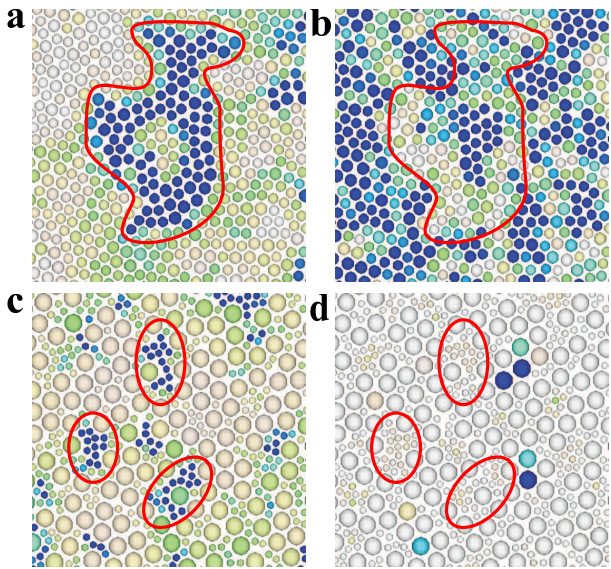}
\end{center}
\caption{(Color online.) (a) and (b) are snapshots of the
system at state B, $\phi=0.66$; (c) and (d) are at state E,
$\phi=0.66$.
The simulation box is four times
as large as those snapshots in each side.  In (a) and (c),
particles are colored based on mobility $\Delta r_j^2$, with
darker colors indicating more mobile particles.  The darkest
color corresponds to $\Delta r_j^2$=0.25$\sigma_s^2$.  
The displacement time scale $\Delta t$ is chosen to maximize the
non-Gaussian parameter, and is $\Delta t= 5 \times 10^4$ for (a)
and $2 \times 10^4$ for (c).
In (b) and
(d), particles are colored based on the hexagonal order parameter
$\psi_6^j$, with the darkest color corresponding to
$\psi_6^j$=0.8.  The outlined regions are guides to the eye.
}
\label{snapshotfig}
\end{figure}

Cooperative motion of groups of particles is a common
phenomenon as the glass transition is approached
\cite{Ediger2,Eric1,Harrowell,Berthier,Doliwa,Eric2}, and
this behavior is thought to be more common in fragile glasses.
Figure \ref{snapshotfig}(a) and (c) show snapshots of the systems
at $\phi$ = 0.66 at state B and state E, where particles are
colored based on $\Delta r_j^2$, and groups of highly mobile
particles are seen (darker colors).  To define displacements, here
(and for the subsequent analysis in this work) we focus on the
time scale $\Delta t^*$ for which the non-Gaussian parameter is the
largest.  The non-Gaussian parameter is defined as 
\begin{equation}
\alpha_2(\Delta t) = \frac{3 \langle \Delta r^4 \rangle}{5 \langle
\Delta r^2 \rangle^2} - 1 
\end{equation}
with displacements $\Delta r$ measured over the lag time $\Delta
t$, and the factor of 3/5 chosen so that $\alpha_2 = 0$ for
a Gaussian distribution \cite{rahman64}.  $\alpha_2$ is larger
when the tails of the distribution become broad.
Prior work identified the time scale $\Delta
t^*$ that maximizes $\alpha_2$ as related to cage rearrangements
\cite{kob97,donati98,Eric1}.  In our current work, we computed
$\alpha_2$ separately for the large and small particles, finding
similar values for $\Delta t^*$.  For our analysis, we will use
$\Delta t^*$ based on $\alpha_2$ calculated for the small particles,
and our results are not sensitive to this choice.

Examining our data, we observe clusters with cooperative
motion in our systems, some of which are circled in
Fig.~\ref{snapshotfig}(a,c).  To look for the connection between
the cooperative motion and fragility in our sample, we need to
characterize the cooperative motion.  We compute a correlation
function described as $S(R)=\langle \Delta \vec{r}_i \cdot \Delta
\vec{r}_j \rangle/\langle |\Delta \vec{r}|^2 \rangle$, where $R$ is
the distance between particles $i$ and $j$, and $\Delta \vec{r}_i$
is the displacement of particle $i$ at $\Delta t$=$\Delta t^*$
\cite{Doliwa}.  We find that $S(R)$ shows exponential decay with
$R$, which was previously observed in experiments \cite{Eric2}
and simulations \cite{Doliwa}.  This exponential decay yields a
decay length $\xi$, which we plot in Fig.~\ref{phidependence}(c)
as a function of $\phi$ for our six representative states.
While all samples have similar short ranged cooperative motion
at $\phi \approx 0.60$, we see a variety of behavior as the glass
transition is approached.

Figure \ref{phidependence}(c) shows little relationship with the
magnitude of $\xi$ and the fragility, so we focus on the growth
rate of $\xi$, $\partial \xi/\partial \phi$.  This is sensible
since the fragility index $D$ relates to the growth of
$\tau_\alpha$ as $\phi$ increases.  
Figure \ref{contourfig}(c) shows the contour plot of $\partial
\xi/\partial \phi$ in a ($\sigma_L/\sigma_S$, $\phi_L/\phi_S$)
plane.  This plot has a rough qualitative similarity to the
fragility [Fig.~\ref{contourfig}(a)]. 
Figure~\ref{fragilityfig}(b) more directly shows
that $D$ is related to $\partial \xi/\partial \phi$, with
distinct behaviors for $\sigma_L/\sigma_S < 1.4$ systems and
$\sigma_L/\sigma_S > 1.4$ systems.  In $\sigma_L/\sigma_S < 1.4$ (triangles),
fragile liquids have larger increase of $\xi$ with respect to
$\phi$ and it is natural that the increase of $\tau_\alpha$ is
related to this.  This is also suggested by experimental studies
of colloidal suspensions \cite{Eric1,Berthier,Eric2}.  In contrast,
the opposite relationship is seen for the large size ratio states
[circles in Fig.~\ref{fragilityfig}(b)].  The most fragile states
(small $D$) correspond with those where $\xi$ grows least as the
glass transition is approached.  This unusual behavior seems to be
a key result for understanding the fragility of large size ratio
binary mixtures.

To further understand the relation between $\xi$ and fragility in
large size ratio systems, we consider the dynamical difference
between the two particle species.  We compute the mean square
displacement $\langle \Delta r^2 \rangle$ of large and small
species separately at a variety of states, and find that the large
particles are always significantly slower than the small particles.
We next consider how the motion of small and large particles is
coupled.  To quantify this, we compute the correlation between the
directions of motion of a large particle and its neighboring small
particles as $\Theta = \langle \frac{1}{n_i} \sum_i \cos\theta_{ij}
\rangle$ where $j$ indicates a large particle, $i$ are the nearest
neighbor particles for particle $j$, $n_i$ is the number of these
neighbors, $\theta_{ij}$ is the angle between $\Delta \vec{r}_i$
and $\Delta \vec{r}_j$, $\Delta \vec{r}_j$ is the displacement of
$j$ particle at $\Delta t$=$\Delta t^*$, and the angle brackets
indicate a time average and an average over all large particles
$j$.  $\Theta=1$ indicates that particles around a large particle
move in the same direction as the large particle on this time
scale, while $\Theta=0$ means that their movements are
uncorrelated with the large particle.
Figure \ref{contourfig}(d) shows the contour plot of $\Theta$ in
the plane of ($\sigma_L/\sigma_S$, $\phi_L/\phi_S$).  $\Theta$
ranges from 0.7 (upper left corner) to 0.4 (lower right corner).
Cooperative motion decreases as the size ratio increases, and as
the number of large particles decreases.

If large particles move slower and small particles move
independently of the large ones [the lighter-shaded region in
Fig.~\ref{contourfig}(d)], this suggests that the large particles act
as slow-moving walls within the large size ratio systems
\cite{moreno06,voigtmann09}.  The small
particles are trapped in pores between the large
particles [see Fig.~\ref{snapshotfig}(c)].  In confined geometries,
$\tau_\alpha$ can dramatically change compared to an unconfined
system at the same temperature and density \cite{Kob3,Nugent,Kim}.
Confined systems with free surfaces typically have smaller values
of $\tau_\alpha$, while those with rigid walls have larger values
of $\tau_\alpha$ \cite{mckenna05}.  Our data suggest the latter
situation is relevant for our large size ratio systems, that
crowding due to the large particles slows the motion of
the small particles.  This has been seen in prior work
\cite{moreno06,voigtmann09} and here we examine this behavior in
more detail.

Here, we suggest the fragility of large size ratio systems
may be connected to the ``strength'' of confinement effects.
Less fragile liquids may have more mobile walls, such as state C
with a large value of $\Theta$ (implying small and large particles
move together).  Or, less fragile liquids may have a larger spacing
between the large particles, such as state F, with a relatively
small value of $\phi_L / \phi_S$; here, they are less confined.
In contrast, the more fragile states D and E have small values
of $\Theta$ and smaller distances between large particles.
These systems are ultra-confined, where the spacing
between large particles is of order $\xi$ or even smaller, thus
limiting $\partial \xi / \partial \phi$.

\subsection{Influences of particle mobility at constant $\phi$}

We also investigate the relationship between local structure and
local mobility in our binary systems.  According to prior work,
the mobility of particles decreases when the particles are in
hexagonally ordered regions (for 2D simulations) \cite{KAT,Tarjus},
which is also hinted at in 3D colloidal experiments
\cite{weeks02}.
Thus, we focus on the spatial distribution of mobility and
hexagonal structure.  
Figure \ref{snapshotfig}(a) shows a snapshot of the system in state
B at $\phi$=0.66 with the darker colors indicating particles
with larger values of $\Delta r_j^2$, and Fig.~\ref{snapshotfig}(b)
shows the same snapshot coloring the particles by their value
of $\psi_6^j$.  The circled regions in Fig.~\ref{snapshotfig}(a,b)
show that mobile regions correspond to regions with less ordering.
We calculate the Pearson correlation coefficient $C(\Delta r_j^2,
\psi_6^j)$, finding $C=-0.13$, supporting the idea that mobility
is slightly anticorrelated with hexagonal ordering, consistent
with prior work.

For large size ratio systems, we see little hexagonal structure on
average [curves E and F in Fig.~\ref{phidependence}(b)],
but this does not preclude the possibility that locally there may be
hexagonal ordering which influences the dynamics.  To check
this, we compare the local mobility with local structure.  
Figure \ref{snapshotfig}(c) and (d) show $\Delta
r_j^2$ and $\psi_6^j$ for state E, with a much larger size ratio,
and here there is no correspondence ($C=0.01$).  Overall, for the
large size ratio samples ($\sigma_L/\sigma_S > 1.4$), we never
observe any ordered structures at any area fraction.  However,
we cannot rule out the possibility that there is subtle
ordering that might be present and influencing the dynamics.

\begin{figure}
\begin{center}
\includegraphics[width=6cm]{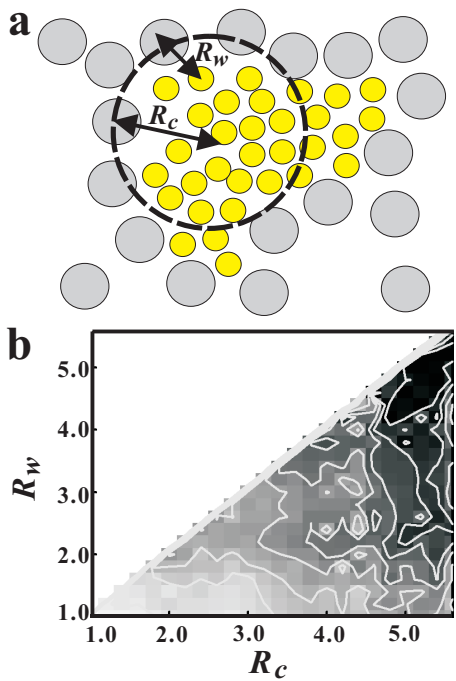}
\end{center}
\caption{(Color online.)  
(a) Schematic showing a group of small particles temporarily
confined within a region bounded by large particles.  Each small
particle is assigned a value $R_w$ which is its closest distance
to any large particle (the large particles act as ``walls'').
Within each region, the local maximum of $R_w$ identifies the
``confinement pore size'' $R_c$ for that region.  Thus, $R_c$ is
the maximum radius of a circle that fits within the bounded region.
(b) A contour plot for the mean square displacement $\langle \Delta
r^2 \rangle$ (at fixed $\Delta t= \Delta t^*$)
as a function of the effective pore size $R_c$ where the given
particle is located, and the distance of that particle
from the nearest ``wall'' $R_w$,
for state E with $\phi=0.66$.  Darker shading corresponds with
high mobility.  The motion of particles is faster in larger pores
(large $R_c$) and when further from the walls (larger $R_w$ for
a constant $R_c$).  The length scales are
in terms of the small particle diameter.  For these data, $R_c^* = 3.6$; only
10\% of regions have $R_c > R_c^*$.
}
\label{mobility}
\end{figure}

Next, we consider the relationship between the confinement effects
and mobility in large size ratio systems.  The confinement effects
are composed of both confinement size effects and confinement
surface effects.  For confined colloidal suspensions, particle
motion was slower in more confined spaces, but there was not a
strong influence from the confining surfaces \cite{Nugent}.
Numerical simulations show that the particles
move slowly near rough walls, and quickly near smooth walls
\cite{Kob3}.  We thus wish to distinguish between finite size
effects and interfacial effects.  First, we define clusters of large
particles as those large particles separated by a distance less
than $1.4 \sigma_L$, and these clusters form ``walls''
surrounding small particles.  In some cases, a connected cluster
of large particles completely surrounds a group of small
particles, as sketched in Fig.~\ref{mobility}.  Within such a
region, we compute the distance $R_w$ from each small particle to the
nearest ``wall'' particle.  The maximum value of $R_w$ within the
confined region defines $R_c$, the effective ``confinement pore
size.''  $R_w$ and $R_c$ are indicated in Fig.~\ref{mobility}(a).
$R_w$ is calculated per small particle, and $R_c$ per pore;
both of these are functions of time.  The dependence of specific particles'
behaviors on $R_w$ should give insight into interfacial effects,
and the dependence of pore-averaged behavior on $R_c$ should give a separate
insight into finite size effects, although of course these two effects
are likely both present simultaneously.

Figure~\ref{mobility}(b) shows a contour plot of $\langle r^2
\rangle$ as a function of a ($R_w$, $R_c$) plane at state E and
$\phi$ = 0.66.  The results are located only at the lower right of
the graph as $R_c \ge R_w$ from our definitions.  The darker region at
the upper right corresponds to high $\langle r^2 \rangle$, showing
that the mobility increases inside large pores (large $R_c$) and
far from walls (large $R_w$).  Given the incommensurate sizes of
the large and small particles, it is difficulty for small particles
to pack well near the ``walls'' [see Fig.~\ref{snapshotfig}(c)] and
so not surprisingly our results are consistent with simulated rough
walls \cite{Kob3}.  There is also a slight gradient of increasing mobility
as a function of pore size $R_c$ for fixed $R_w < 1.5$, indicating that there
is a finite size effect in addition to an interfacial effect.  That is, smaller
pores (smaller $R_c$) have more particles close to the pore walls, and thus
experience stronger interfacial effects, but the data indicate that the
influence of the interface on adjacent particles is less within large pores.

\begin{figure}
\begin{center}
\includegraphics[width=6cm]{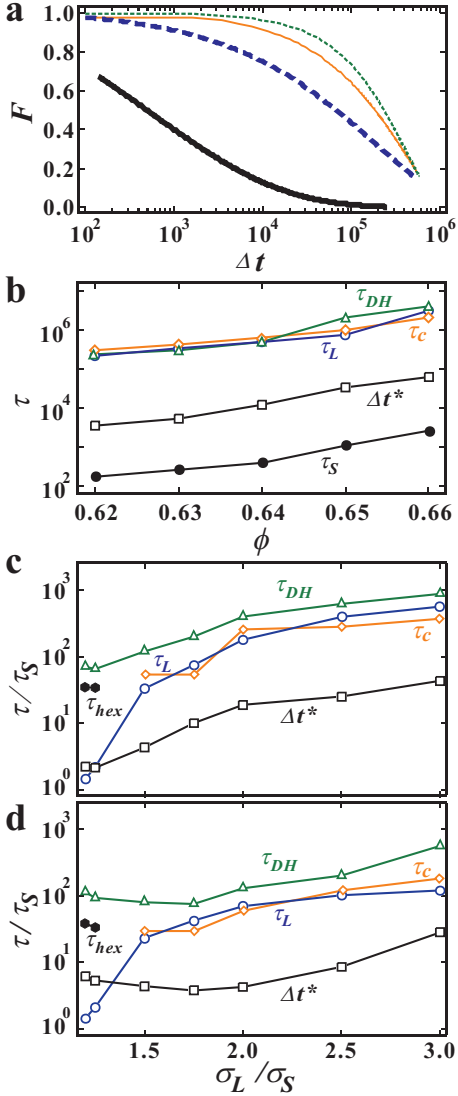}
\end{center}
\caption{(Color online.)  (a) The time dependence of $F(k,\Delta t)$
for small particles (thick solid (black) line; $k = 2 \pi/\sigma_S$)
and for large particles (thick dashed (blue) line; $k =
2 \pi/\sigma_L$) at state E and $\phi$ = 0.66.  
The thin solid (red) line indicates the breakup of the largest
confined regions, and the thin dashed (green) line indicates when
the most mobile particles become less mobile.  (b) The typical
structural relaxation time $\tau_S$ for small particles (filled circles),
the structural relaxation time $\tau_L$ for large particles
(open circles), the confinement lifetime $\tau_c$ (diamonds), 
the dynamical heterogeneity lifetime $\tau_{DH}$ (triangles), 
and the peak time for the non-Gaussian parameter $\Delta t^*$ 
(squares).  
The data are for state E. 
(c) Five time scales ($\tau_L$, $\tau_c$, $\tau_{DH}$, 
$\Delta t^*$, and $\tau_{hex}$) normalized by $\tau_S$ as a function of 
$\sigma_L/\sigma_S$ at $\phi_L/\phi_S$ = 2.0 at $\phi$ = 0.66.
$\tau_{hex}$ is the lifetime of hexagonal order and corresponds to the filled
hexagon symbols.
(These data go through state points A, C, and E; see Fig.~\ref{contourfig}.)  
(d) Five time scales ($\tau_L$, $\tau_c$, $\tau_{DH}$,  
$\Delta t^*$, and $\tau_{hex}$) normalized by $\tau_S$ as a function of 
$\sigma_L/\sigma_S$ at $\phi_L/\phi_S$ = 0.5 at $\phi$ = 0.66. 
(These data go through state points B, D, and F.)
There are only a few confinement regions when
$\sigma_L/\sigma_S < 1.5$, preventing a clear determination of $\tau_c$,
so those curves are clipped in (c) and (d).
}
\label{lifetime}
\end{figure}

To obtain further evidence for the relationship between
the confinement effects and dynamical heterogeneity, we also
investigate the temporal relationship between local structures and
dynamical heterogeneity.  We calculate the intermediate function
$F(k,\Delta t)$ for small and large particles separately at fixed
$k$ = 2$\pi/\sigma_S$ for small particles (thick solid (black)
line in Fig.~\ref{lifetime}(a)) and $k$ = 2$\pi/\sigma_L$ for large
particles (thick dashed (blue) line in Fig.~\ref{lifetime}(a)).
The structural relaxation time scales for small and large
particles, $\tau_S$ and $\tau_L$, are set by $F(k,\tau)$
= 1/e.  We next consider all of the confined regions and the
distribution of region sizes $R_c$.  We determine the distribution of
all pore sizes $R_c$ (taken over the entire simulation run), and
find the threshold size $R_c^*$ for the top 10\% of this distribution.
For each pore at each time, we define $W_c(t)=1$ if $R_c(t) > R_c^*$
and $W_c(t)=0$ otherwise.  Typical values of $R_c$ range from 2.8
to 5.6.
The temporal correlations
of the regions are given by $\langle W_c(\Delta t)\cdot W_c(0)
\rangle/\langle W_c(0)^2 \rangle$, plotted as the thin solid (red)
line in Fig.~\ref{lifetime}(a).  The typical lifetime of large
regions is given as $\tau_c$ where the correlation drops to $1/e$.
Similarly, for each particle we define a parameter $W_{DH}$ which
is equal to 1 if the particle's displacement $\Delta r$ at that
time is within the top 10\% of the displacement distribution, and 0
otherwise.  The correlation $\langle W_{DH}(\Delta t)\cdot W_{DH}(0)
\rangle/\langle W_{DH}(0)^2 \rangle$ is plotted as the thin dashed
(green) line in Fig.~\ref{lifetime}(a), and $\tau_{DH}$ is defined
by the $1/e$ time again; this is 
the time scale over which particles
exchange between being fast and slow, as mentioned in the
Introduction.

Figure \ref{lifetime}(b) shows the $\phi$ dependence of these
four time scales ($\tau_S$, $\tau_L$, $\tau_c$, $\tau_{DH}$
and $\Delta t^*$).  The fastest time scale is $\tau_S$ (filled
circles), and the large particles are
much slower ($\tau_L$, open circles). What is more
notable is that $\tau_L \approx \tau_c$, in other words, large
particle motions relate to the relaxation of confined regions.
This is further evidence that the large particles form walls.
Furthermore, $\tau_c \approx \tau_{DH}$ is observed, connecting the
confinement-induced dynamics with the dynamically fast particles.
Faster particles exchange identities with slower particles when
the confining walls rearrange.

If the spatial dynamical heterogeneity is actually induced by the
confinement effect, we would expect to see $\tau_L \approx \tau_c
\approx \tau_{DH}$, and the correspondence should be strongest
for large size ratio systems.  Figure \ref{lifetime}(c) and (d)
show these time scales normalized by $\tau_S$ as a function of
the size ratio $\sigma_L/\sigma_S$  at $\phi_L/\phi_S$ = 2.0
and 0.5, respectively; the data are for $\phi=0.66$.  Indeed,
for $\sigma_L/\sigma_S \ge 1.5$, we find $\tau_L$ is similar
to $\tau_c$ and $\tau_{DH}$.  Not surprisingly, the relaxation
time scale $\tau_c$ for the confinement effect is governed by the
relaxation time $\tau_L$ for the large particles which define the
pores; more significantly, the lifetime of dynamically heterogeneous
regions is also connected to the behavior of the large particles.

On the other hand, we find $\tau_L/\tau_S \rightarrow 1$ when
$\sigma_L/\sigma_S \rightarrow 1$, which is to be expected.
In these cases, $\tau_{DH}$ remains large, showing that
here the dynamical heterogeneity is less influenced by the
relaxation of the large particles.  However, 
$\tau_{DH}$ is same order as the lifetime of hexagonal order,
as seen in Fig.~\ref{lifetime}(c,d) by comparing the triangles
($\tau_{DH}$) with the filled hexagons ($\tau_{hex}$).  
This new time scale, $\tau_{hex}$, is defined in a similar way as
the other time scales:  particles with $\psi_6>0.8$ are
considered hexagonal, and the correlation time for having
$\psi_6>0.8$ is $\tau_{hex}$.  This time scale is only relevant
for samples with reasonable amounts of hexagonal order [compare
Fig.~\ref{snapshotfig}(b,d)], and so is only shown for samples
with small size ratios.  (This is why it is not shown in
Fig.~\ref{lifetime}(b) which has $\sigma_L/\sigma_S = 2.5$.)
It is further evidence
that local hexagonal order influences the dynamics in similar
size ratio systems.  In these cases, the confinement effect is
more likely due to hexagonal regions composed of both particle
species, rather than networks formed by only the large particles,
and thus $\tau_c$ becomes ill-defined and we do not show it in
Fig.~\ref{lifetime}(c,d).

The competition between hexagonal ordering and crowding due to 
large particles likely accounts for the cases in
Fig.~\ref{lifetime}(c,d) where $\tau_{DH} > \tau_L$.  For example,
states with $\sigma_L/\sigma_S = 1.5$ still have regions of
hexagonal ordering composed of both particle sizes.  In these cases
$\Theta$ is small [see Fig.~\ref{contourfig}(d)] and the confinement
effect is weak.  Another example where $\tau_{DH} > \tau_L$
is the state with $\sigma_L/\sigma_S = 3.0$, $\phi_L/\phi_S=0.5$
[Fig.~\ref{lifetime}(d)].  The large particles are scarce, but
result in a strong confinement influence; however, the more numerous
small particles can themselves form hexagonal patches, influencing
their mobility.  We expect that for 3D glass-formers, local
crystalline order is much less significant, and so confinement
effects would more strongly determine $\tau_{DH}$ in all cases.

Furthermore, we consider the relationship among $\Delta t^*$
and other time scales (Fig.~\ref{lifetime}(c) and (d)).  As a
reminder, $\Delta t^*$ is the time scale at which the small particle
displacement distribution is the most non-Gaussian.  This time
scale is used to define the mobility of particles and so is part
of the definition of a ``mobile'' particle and thus $\tau_{DH}$.
$\Delta t^* > \tau_S$
shows that small particles move appreciable distances during the
time $\Delta t^*$.  That is, the non-Gaussian displacements are
over significant distances, enough to relax the small particle
structure, and this becomes more true at larger
size ratios, although this motion is all localized within a region
defined by nearby large particles.  
In all cases $\tau_{DH} > \Delta t^*$ showing that slow and fast
regions do not change identities with each particle rearrangement,
but rather take longer times to change.  Particles may rearrange
several times confined within a large pore (several $\Delta t^*$)
before the pore rearranges and the particles change their mobility.
An important caveat is that these observations are for
the average behavior of particles, and we are not implying that
the connections between pore sizes and dynamics are strongly
deterministic.  Nonetheless the hierarchy of time scales in
Fig.~\ref{lifetime} is suggestive of nontrivial connections between
the spatial arrangements of the large particles and the long-lived
lifetimes of dynamically unusual regions, both fast and slow.

\subsection{Aging}

\begin{figure}
\begin{center}
\includegraphics[width=7.8cm]{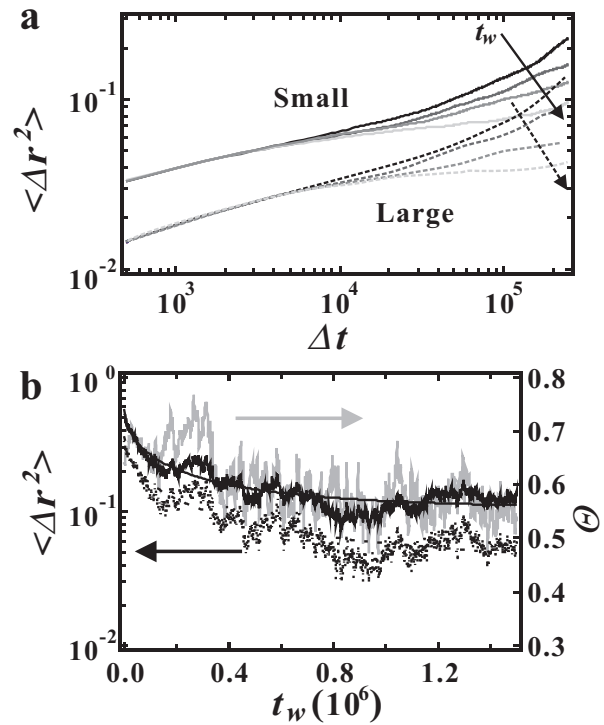}
\end{center}
\caption{(a) $\langle \Delta r^2 \rangle$ as a
function of lag time $\Delta t$ for small particles (solid lines)
and for large
particles (dashed lines) at $t_w = 10^3, 10^4, 10^5$ and $10^6$
at state E and $\phi$ = 0.72.  The mobility of both particle
species decreases with respect to aging time.  (b) The dark black
lines correspond to $\langle \Delta r^2 \rangle$ at $\Delta t =
10^5$ as a function of aging time $t_w$ for small particles (solid)
and large particles (dashed).  The smooth curves are
fits with a stretched
exponential, and we find that the decay time of large particles is
slightly shorter than that of small particles.  The gray line
show the waiting time dependence of
$\Theta$, the correlation between the displacements of a large
particle and its surrounding small particles.  The choice of $\Delta t=10^5$ is
arbitrary although panel (a) shows that this is a reasonable choice to capture
the slowing dynamics.
}
\label{aging1}
\end{figure}

Next we investigate aging dynamics in binary samples with large
size ratios.  We study state E at $\phi = 0.72$
where it is in a glassy state ($\phi_g \sim 0.68$ for this state).
Figure \ref{aging1}(a) shows $\langle \Delta r^2 \rangle$ for
small and large species separately at waiting times $t_w = 10^3,
10^4, 10^5$ and $10^6$.  At short time scales ($\Delta t < 5000$),
$\langle \Delta r^2 \rangle$ is independent of $t_w$.  Particles
move within cages formed by their nearest neighbor particles.
At longer time scales $\Delta t > 5000$, $\langle \Delta r^2
\rangle$ increases as the cages rearrange and allow particles to
move to new positions.  This upturn in $\langle \Delta r^2 \rangle$
decreases with increasing $t_w$, indicating the aging of the
system, similar to what has been seen in experiments
\cite{courtland03,Gianguido, Jenn}.  Figure~\ref{aging1}(b) shows
the $t_w$ dependence of $\langle \Delta r^2 \rangle$ at fixed
$\Delta t$ = 10$^5$.  Though the results at small $t_w$ ($t_w <
\Delta t$) are hard to interpret as the dynamics change during $\Delta
t$, we can observe the clear temporal change of $\langle \Delta
r^2 \rangle$ for both particle species.  For ``old'' systems, the plateau
of $\langle \Delta r^2 \rangle$ extends over a large range of time scales, 
with the plateau height
corresponding to the
cage size.  Within our uncertainty, the large and small particles
age almost at the same rates, as suggested by the similar shapes
of $\langle \Delta r^2 \rangle$ curves (Fig.~\ref{aging1}(a)).
Again, these observations are consistent with experiments in binary
colloidal glasses \cite{Jenn}.

\begin{figure}
\begin{center}
\includegraphics[width=6cm]{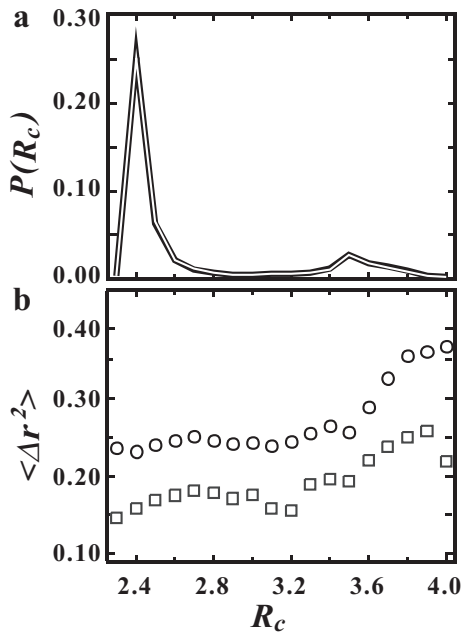}
\end{center}
\caption{(a) The probability distribution $P(R_c)$ of confinement
pore sizes at $t_w=10^3$ (black line) and $t_w=10^6$
(superimposed white line), showing the distributions are
essentially identical.  This is for state E with $\phi=0.72$.
(b) Mean squared displacements
$\langle \Delta r^2 \rangle$ for fixed $\Delta t = 10^5$ as a function of
the confinement pore size $R_c$ at $t_w = 10^3$ (circles)
and $t_w = 10^6$ (squares), for the same state as (a).  The
upward trend (faster motion for larger pores) is similar to what
is seen in equilibrated liquids [Fig.~\ref{mobility}(b)].
For these data $\Delta t = 10^5$ to capture time scales over which
the dynamics age [see Fig.~\ref{aging1}(b)].
}
\label{aging2}
\end{figure}

We wish to know a reason for the slowing dynamics with respect to
the waiting time.  First, we confirm that the overall structure is
unchanged with age:  we compute the pair correlation function $g(R)$
at $t_w$ = 1000 and 10$^6$ and cannot observe any difference between
them, similar to prior observations in simulations \cite{kob00}
and experiments \cite{Gianguido,Jenn}.  $g(R)$ is a spatial
average over the whole system, so we next consider the relation
between local structure and local mobility.  We use the confinement
pore size $R_c$ to characterize the local structure of the particles
within a given confining region.
Figure \ref{aging2} shows $\langle \Delta r^2 \rangle$
and the probability distribution $P(R_c)$, both as functions of
$R_c$ at $t_w$ = 1000 and 10$^6$.  We find that $P(R_c)$ does not
change at all and this is further evidence that the structure does
not change.  However, the mobility decreases with age
with the dependence on $R_c$ relatively unchanged
other than the amplitude.  This implies that the mobility
decrease is not due to local structural changes, but rather an
average slowing of the whole system \cite{Gianguido}.

\begin{figure}
\begin{center}
\includegraphics[width=6cm]{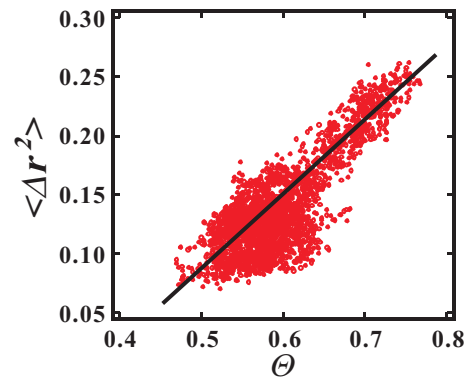}
\end{center}
\caption{
(Color online.) Scatter plot of the mean square displacement
$\langle \Delta r^2 \rangle$ as a function of $\Theta$ at state
E and $\phi$ = 0.72,
using $\Delta t=10^5$.  Each
point corresponds to a different $t_w$, the time since
the start of the simulation.  The solid line is a least squares
fit.  The data indicate that mobility is linked to the correlation
of the motion between large particles and their nearest neighbors.
}
\label{aging3}
\end{figure}

Next, we focus on the importance of confinement which strongly
influences the dynamics of our equilibrated
large size ratio binary systems.
As noted above, the confinement strength depends on the confinement
size $R_c$ and the effective rigidity of the walls, that is, $\Theta$.  
In aging systems,
$P(R_c)$ does not change with respect to $t_w$, but $\Theta$
could depend on $t_w$ since $\Theta$ is a dynamical property rather
than a structural property.  
We compute $\Theta(t_w)$ at fixed $\Delta t$ = 10 $^5$,
shown as the gray line in Fig.~\ref{aging1}(b).  The behavior of $\Theta$ looks similar
to the mobility change of both particles.  Figure \ref{aging3}
shows a scatter plot of the mean square displacement $\langle
\Delta r^2 \rangle$  as a function of $\Theta$, for
all large particles and all waiting times $t_w$.  We can clearly see
the correlation between the mobility and $\Theta$.  When $\Theta$
decreases, the cooperative motion between small and large particles
is less, in other words, the rigidness of walls increases.  This
result implies that
confinement effects become stronger during
aging and it may help explain the slowing down of the mobility.  However,
we don't know why $\Theta$ decreases as the sample ages.

\section{Conclusion}

We have examined the glass transition in binary mixtures with a
large size ratio, finding results that are distinct from binary
mixtures with smaller size ratios.  Systems with smaller size ratios
are often studied, and the utility of using two particle sizes in
those cases is to frustrate the packing and prevent crystallization.
Crystals are also frustrated in our simulations with large size
ratios, and in addition we find
several new results.  First, we have investigated fragility of
binary systems.  The fragility of binary systems with size ratios
close to 1 ($\sigma_L/\sigma_S<1.4$) is related to the growth of
ordered regions:  more fragile liquids show dramatic increases in
hexagonal order as the glass transition is approached \cite{KAT,
Sun, Coslovich}.  However, systems with larger size ratios do
not show this relation.  Both types of systems have fragilities
which are related to the growth of a dynamical length scale $\xi$,
although the sign of this correlation is opposite for the small
and large size ratio systems.

The data show that in large size ratio systems, large particles
act as quasi-immobile walls which confine the small particles
and slow the dynamics overall.  
The large particles define regions with a
range of sizes.  Small particles in large regions find it easier
to move, even if within that region they may be adjacent to a
large particle at the boundary of the region.  Small particles in
smaller regions are much less mobile.  In addition to these
finite size effects, there are interfacial effects:
small particles near large particles move slower than those
farther away.  Those results are what is often seen in experiments
and simulations of confined supercooled liquids.  Furthermore,
as is often observed in simulations and experiments, we find some
particles are unusually mobile.  Our new observation is that the
length of time which these particles stay unusually mobile is
connected to the lifetime of the confinement effect and thus the
relaxation time of the large particles.

Finally, we also investigate aging dynamics in large size ratio
systems.  Below the glass transition, the mobility decreases with
respect to the waiting time, though we cannot observe any structural
change.  We find that in ``younger'' systems that the motion of
large particles is correlated with the motion of their neighbors,
but that in ``older'' systems this correlation is markedly smaller.
This correlation (or lack of it) relates to the confinement effect,
suggesting that the large particles become more rigid confiners
in older samples.

It is important to note that simulations of softer particles with
a charged (Yukawa) potential find results different from ours,
pointing out that our results are not completely generalizable.
In simulations with a size ratio $1 : 5$, the large particles
crystallized \cite{kikuchi07}.  In these cases, the large particles
did not rearrange but rather moved on their lattice sites, and small
particles could only move by diffusive hopping motions between
the crystal interstices.  This is in contrast to our simulations
where the large particles are always able to rearrange (albeit
more slowly than the small particles).

Overall, our results suggest that in binary soft sphere systems, the
effect of the large particles to induce finite size effects within
the sample play an interesting role in the dynamics.  Two relevant
variables are the finite sizes of regions between large particles,
and the effective rigidity of the large particles.  While our
simulation studies 2D systems, the results are similar to prior
observations in 3D binary colloidal experiments \cite{Jenn,Narumi}
with moderate size ratios.  While the effects are easiest to see
with large size ratios, one implication of our simulations is
that these effects may be relevant although less obvious in binary
systems of smaller size ratios.  Indeed, one of the goals of our
simulations was to understand the effect of structure by using systems
where structural heterogeneity is more obvious.  These results
may also have implications for studies of nanocomposites, where
inclusions into polymer glasses can dramatically affect the
properties of materials \cite{Rittigstein,Kropka}.

\section*{Acknowledgments}
This work was supported by the National Science Foundation under
Grant No. DMR-0804174.  R. K. was supported by the Japan Society
for the Promotion of Science.

\end{document}